\documentclass[11pt]{article}

\usepackage[final]{acl}

\usepackage{times}
\usepackage{latexsym}

\usepackage[T1]{fontenc}

\usepackage[utf8]{inputenc}

\usepackage{microtype}

\usepackage{inconsolata}

\usepackage{graphicx}
\usepackage{amsmath}
\usepackage{booktabs}
\usepackage{amsfonts}

\usepackage[most]{tcolorbox}
\usepackage{xcolor}

\definecolor{baselineorange}{RGB}{220,100,20}
\definecolor{oursgreen}{RGB}{40,140,70}

\usepackage[most]{tcolorbox}
\usepackage{enumitem}

\title{Noise Adaptive Streaming Audio-Visual Speech Token Enhancement for Robust Full-Duplex Spoken Dialogue Models}

\author{
  \textbf{Bella Godiva},
  \textbf{Yeonju Kim},
  \textbf{Yong Man Ro\thanks{Corresponding author.}}
\\
  Integrated Vision and Language Lab, KAIST, South Korea
\\
  \small{
    \texttt{\{bellagodiva, yeonju7.kim, ymro\}@kaist.ac.kr}
  }
}

\begin{document}
\maketitle

\begin{abstract}
Full-duplex spoken dialogue systems enable simultaneous listening and speaking, but their audio-only perception often fails under background noise and overlapping speech, leading to incoherent responses. Recent audio-visual dialogue approaches show that incorporating visual cues such as lip movements improve robustness under audio corruption. However, existing approaches often adapt the large speech dialogue model itself
to process visual input, requiring costly multimodal training. We propose AV-STE, a modular streaming audio-visual front-end that restores corrupted semantic speech tokens from noisy audio and lip video before they reach the speech LLM. The downstream dialogue model remains entirely frozen, preserving its pretrained conversational capabilities. When integrated with frozen Moshi, AV-STE improves average GPT-4o-judged response coherence from 1.42 to 1.91 under same-dataset speaker interference while largely preserving turn-taking behavior. Gains also transfer to out-of-domain Seamless Interaction. Code and models are available: \url{https://github.com/bellagodiva/av-ste.git}
\end{abstract}

\section{Introduction}
\label{sec:intro}
Full-duplex spoken dialogue models enable real-time interaction by listening and speaking simultaneously, supporting natural behaviors such as
interruptions and backchannels. However, their robustness to acoustic
interference remains underexplored. In real-world environments, background
noise and competing speakers can corrupt the user's speech representation,
leading to degraded understanding and less coherent responses.

\begin{figure}[t]
\centering
\includegraphics[width=\linewidth]{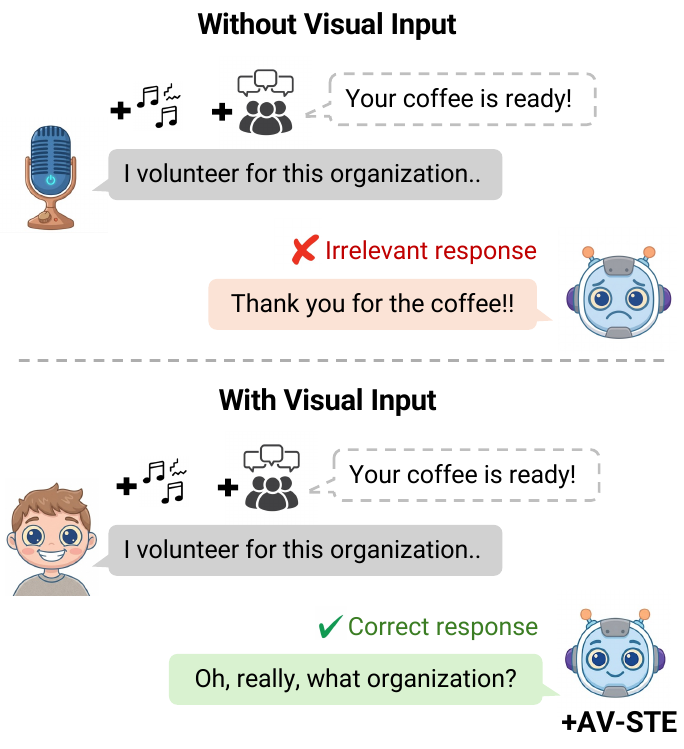}
\caption{In noisy environments, background sounds such as music, crowd noise, or interfering speech can degrade speech perception in full-duplex spoken dialogue models, leading to irrelevant responses. AV-STE uses visual lip cue to enhance corrupted speech token and improve response coherence under noisy conditions, without retraining the underlying speech LLM.}
\label{fig:problem_definition}
\end{figure}

Since current full-duplex dialogue models depend primarily on audio input, their performance can degrade substantially under noisy conditions. In human communication, visual cues such as lip movements provide complementary information when speech is ambiguous or corrupted. Prior work in audio-visual speech recognition (AVSR)~\citep{auto_avsr,mms-llama} has demonstrated the benefit of visual information for robust speech understanding. However, AVSR systems primarily predict text, whereas native speech-to-speech dialogue models operate directly on speech-token representations and must process user input continuously with low latency. Using AVSR would introduce an additional speech-to-text and text-to-speech-token conversion stage. We therefore focus on directly enhancing corrupted speech tokens using visual information.

AV-Dialog~\citep{av-dialog} further demonstrates that visual information can improve spoken dialogue robustness by grounding the dialogue model in
audio-visual input through large-scale training on paired audio-visual
conversational data. However, such training is computationally expensive and data-intensive. Moreover, adapting a pretrained dialogue model to a new
modality requires careful training to retain its existing language and
conversational capabilities. This motivates a practical question: can
visual information improve dialogue robustness without retraining the speech LLM itself?

To this end, we propose AV-STE, a streaming audio-visual semantic token
enhancement framework for robust full-duplex dialogue. Rather than adapting
the dialogue model to process visual input, AV-STE uses a streaming
audio-visual encoder and noise-adaptive fusion to restore corrupted semantic speech tokens from noisy audio and lip video before they reach the dialogue model. The enhanced tokens are streamed directly into the pretrained speech LLM, which remains entirely frozen, avoiding costly audio-visual retraining and enabling plug-and-play integration with existing speech-to-speech dialogue systems.

Experiments on LRS3 and Seamless Interaction show that AV-STE substantially
improves semantic-token recovery under non-speech noise and speaker
interference, including same-dataset multi-speaker interference and
out-of-domain conditions. When integrated with frozen Moshi, AV-STE
consistently improves response coherence across noisy SNR levels while
preserving turn-taking behavior. These results demonstrate that visual speech cues can improve the robustness of full-duplex spoken dialogue by recovering corrupted semantic speech tokens without retraining the underlying speech LLM.

\section{Related Work}
\label{sec:related_works}

\subsection{Full-Duplex Spoken Dialogue Models}
Recent speech LLMs can be broadly categorized as half- or full-duplex depending on whether they can listen and speak simultaneously. Many speech LLMs operate in a half-duplex manner~\citep{zeng2024glm,xu2025qwen2,ding2025kimi}, alternating
between listening and speaking. While effective for turn-based interaction,
this limits natural conversational behaviors such as interruption and
backchanneling.

Full-duplex speech LLMs~\citep{moshi,syncllm,zhang2025omniflatten,freezeomni,
minmo,salmonn-omni} instead process streaming user speech while generating
responses, enabling more natural turn-taking. Some systems use modular
listening and speaking components, such as Freeze-Omni~\citep{freezeomni} and MinMo~\citep{minmo}, whereas models such as Moshi~\citep{moshi} operate directly over speech-token streams within a unified architecture. Our work builds on this latter setting and improves robustness to acoustic corruption while keeping the pretrained speech LLM frozen.

\subsection{Audio-Visual Spoken Dialogue Models}
Recent work has also incorporated visual information directly into spoken
dialogue systems. MoshiVis~\citep{moshivis} extends Moshi with visual input for understanding and discussing image content. In contrast, AV-STE uses continuous lip video as a complementary signal for recovering speech information under acoustic corruption.

AV-Dialog \citep{av-dialog} shares a similar motivation of incorporating user face video into dialogue systems. However, it incorporates audio-visual grounding by training the dialogue model on paired audio-visual conversational data. AV-STE instead introduces visual information at the speech-token level, restoring corrupted semantic tokens before they reach the dialogue model. This allows the underlying speech LLM to remain frozen and enables AV-STE to serve as a plug-and-play robustness module for pretrained speech-to-speech dialogue systems.

\begin{figure*}[t]
\centering
\includegraphics[width=\linewidth]{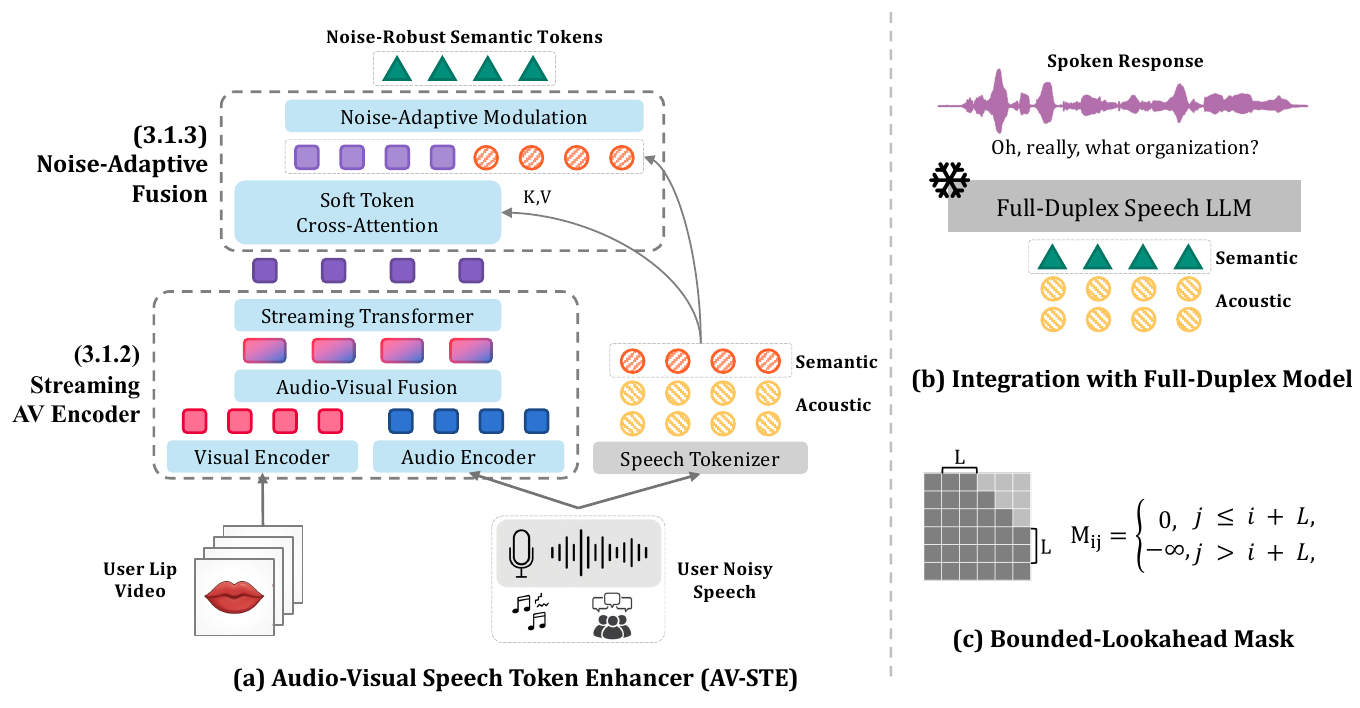}
\caption{
\textbf{Overview of AV-STE}. (a) Given noisy speech and the corresponding lip video, AV-STE first extracts visual speech features using a Streaming Audio-Visual Encoder. The Noise-Adaptive Fusion module then combines these features with speech-tokenizer outputs through soft token cross-attention and noise-adaptive modulation to predict clean semantic tokens. (b) The enhanced semantic tokens are streamed into a frozen full-duplex speech LLM together with the acoustic tokens from the speech tokenizer, enabling more robust response generation under noisy conditions. (c) A bounded-lookahead mask is used to preserve low-latency streaming inference.
}
\label{fig:model_architecture}
\end{figure*}

\subsection{Speech Token Representations}

Recent full-duplex speech dialogue systems operate directly on discrete speech token representations rather than text. Discrete speech representations generally capture different levels of speech
information. Semantic tokens derived from self-supervised speech models such as HuBERT~\citep{hubert} emphasize linguistic content, while neural audio codecs such as EnCodec~\citep{encodec} and SoundStream~\citep{soundstream} encode acoustic information needed for waveform reconstruction.

More recent tokenizers, including SpeechTokenizer~\citep{speechtokenizer} and Mimi~\citep{moshi}, combine semantic and acoustic information within a
multi-codebook representation. In Mimi, the first codebook is semantically
distilled, while the remaining codebooks primarily encode acoustic
information. AV-STE specifically enhances this semantic-token stream while
retaining the acoustic-token streams unchanged, allowing the enhanced
representation to be consumed directly by Moshi. Although we evaluate AV-STE with Mimi and Moshi, the token-level formulation could be adapted to other speech models that use semantic-token representations.
\section{Audio-Visual Speech Token Enhancer (AV-STE)}
\label{sec:method}

\subsection{Overview}
We propose AV-STE, a lightweight audio-visual speech token enhancement module that improves the robustness of frozen full-duplex speech dialogue models under noisy and interfering-speaker conditions. In this work, we focus on semantic speech tokens because they primarily encode linguistic content, which is most critical for speech understanding and coherent response generation. This choice is also motivated by the nature of visual speech, where lip movements provide strong phonetic cues but limited information about fine-grained acoustic attributes such as timbre, prosody, and background noise. We therefore focus on enhancing the semantically distilled token stream while retaining the original acoustic-token streams unchanged.

Given noisy semantic speech tokens $\mathbf{z}^{\mathrm{noisy}}$ and synchronized facial video $\mathbf{v}$ of the target speaker, AV-STE predicts clean semantic speech tokens $\mathbf{y}$ before they are fed into a frozen speech LLM:
\begin{equation}
\mathbf{y}
\sim
p_{\theta}
\left(
\mathbf{y}
\mid
\mathbf{z}^{\mathrm{noisy}},
\mathbf{v}
\right).
\end{equation}
The proposed framework consists of two main components: (1) a \textit{Streaming Audio-Visual Encoder}, which extracts visual speech features from noisy audio and synchronized lip video, and (2) a \textit{Noise-Adaptive Fusion} module, which uses cross-attention to combine visual speech features with noisy tokenizer representations and adaptively adjusts their contribution according to tokenizer uncertainty. The resulting enhanced semantic tokens are then passed directly to the frozen speech LLM, enabling robust audio-visual input processing without modifying the underlying dialogue model. The full architecture of AV-STE is illustrated in Figure~\ref{fig:model_architecture}.

\subsection{Streaming AV Encoder}

We initialize the Streaming AV Encoder from AV-HuBERT \citep{av-hubert}, a self-supervised audio-visual speech model that achieves strong performance on audio-visual speech recognition. AV-HuBERT consists of modality-specific audio and visual encoders followed by a shared Transformer encoder, which extracts synchronized audio-visual features from input speech waveforms and facial video frames at 25 Hz. Since the original AV-HuBERT is bidirectional and designed for offline inference, we modify the backbone to support low-latency streaming inference.

First, we replace all temporal convolutions in the video frontend with causal convolutions using left-only temporal padding, ensuring that each visual feature depends only on current and past frames. Second, we replace bidirectional self-attention in every Transformer layer with causal attention using a bounded-lookahead mask \citep{streaming_asr,emformer}, as illustrated in Figure~\ref{fig:model_architecture}(c). Specifically, we add the following mask to the pre-softmax attention logits:
\begin{equation}
M_{ij}
=
\begin{cases}
0, & j \le i + L, \\
-\infty, & j > i + L,
\end{cases}
\end{equation}
where $L$ denotes the allowed future lookahead in frames. When $L=0$, the model is strictly causal; when $L>0$, the model accesses a bounded number of future frames while preserving streaming inference through fixed-latency future context. A small lookahead improves temporal audio-visual alignment while maintaining low streaming latency. We use $L=4$ (160~ms at 25~Hz) in our main experiments and further ablate different lookahead sizes in Figure~\ref{fig:lookahead}.

To match the target speech token rate used by the downstream speech LLM, we additionally insert a strided causal convolution after the transformer encoder to downsample the feature sequence while preserving the streaming constraint.

\subsection{Noise-Adaptive Fusion}

The streaming AV encoder produces visual speech features $\mathbf{h}^{\mathrm{av}}_t$ that are robust to background noise and interfering speakers. However, visual speech cues can be phoneme-ambiguous, since multiple phonemes may correspond to identical mouth shape \citep{vsp-llm}, making AV encoder predictions less reliable than the original speech tokens under clean and low-noise conditions. In contrast, the speech tokenizer provides fine-grained speech tokens learned from large-scale speech data, but its predictions may degrade under severe noise. To exploit their complementary strengths, we introduce a noise-adaptive fusion module that incorporates tokenizer predictions into the AV representation through cross-attention and adaptively modulates their contribution based on tokenizer uncertainty.

\paragraph{Soft Token Cross Attention. } Directly attending to hard token embeddings can be unreliable under noisy or interfering-speaker conditions, where the tokenizer distribution is often ambiguous. Hard $\arg\max$ selection discards this uncertainty and may provide misleading token-level evidence to the cross-attention module. We therefore attend to the soft tokenizer distribution instead of discrete token embeddings. Specifically, we compute a distribution over tokenizer logits and use it to form a soft token representation:
\begin{equation}
\mathbf{p}_t
    = \mathrm{softmax}\!\left(\frac{\mathbf{l}_t}{\tau}\right),
\qquad
\mathbf{e}_t
    = \mathbf{p}_t\,\mathbf{E}\,\mathbf{W}_{\mathrm{proj}},
\label{eq:soft_lookup}
\end{equation}
where $\mathbf{l}_t \in \mathbb{R}^V$ denotes the tokenizer logits, $\tau$ is a temperature hyperparameter, $\mathbf{E} \in \mathbb{R}^{V \times d_e}$ is a token embedding table learned jointly with the fusion module, and $\mathbf{W}_{\mathrm{proj}} \in \mathbb{R}^{d_e \times d}$ projects the soft embedding to the model dimension. 

We then fuse $\mathbf{e}_t$ with the audio-visual features through cross-attention~\citep{vaswani2017attention}:
\begin{equation}
\hat{\mathbf{c}}_t
    = \mathrm{CrossAttn}\!\left(
        \mathbf{h}^{\mathrm{av}}_t,\;
        \mathbf{e}_{1:T},\;
        \mathbf{e}_{1:T}
      \right),
\end{equation}
\begin{equation}
\mathbf{c}_t
    = \mathrm{LayerNorm}\!\left(
        \mathbf{h}^{\mathrm{av}}_t + \hat{\mathbf{c}}_t
      \right).
\end{equation}
We use AV features $\mathbf{h}^{\mathrm{av}}_t$ as queries and soft token
embeddings $\mathbf{e}_t$ as keys/values, allowing noise-robust visual cues to guide retrieval from the corrupted speech-token representation. The reverse direction performs worse under noise (Appendix~\ref{app:query_ablation}).

To support streaming inference, we apply a bounded-lookahead mask
(Figure~\ref{fig:model_architecture}c) that restricts each query position $t$
to attend only to past key-value positions $t' \leq t + L$, consistent with
the backbone lookahead $L$.

\paragraph{Noise-Adaptive Modulation.}
Although cross-attention incorporates the speech-token representation into the AV representation, the speech-token representation should not contribute equally under all noise conditions. In clean or low-noise speech, the tokenizer predictions are usually reliable. However, in severe noise or interfering speech, the tokenizer distribution may contain incorrect acoustic information. We therefore introduce a frame-level modulation $\lambda_t \in [0,1]$ that adaptively controls how much weight the final representation places on the speech-token representation $\mathbf{e}_t$ versus the cross-attended AV representation $\mathbf{c}_t$ depending on the noise level.

We use the normalized entropy of the tokenizer distribution as a reliability prior, since entropy measures uncertainty over the full token distribution~\citep{laptev2023fast}:

\begin{equation}
\begin{aligned}
\hat{H}_t
    &= \frac{-\sum_{v=1}^{V} p_{t,v} \log p_{t,v}}{\log V}, \\
\hat{\lambda}^{\mathrm{ent}}_t
    &= 1 - \hat{H}_t .
\end{aligned}
\label{eq:entropy_gate}
\end{equation}

where $\hat{H}_t \in [0,1]$ denotes the normalized entropy. A lower entropy indicates a sharper tokenizer distribution and thus higher estimated reliability, while a higher entropy indicates greater uncertainty.

Because entropy alone may not capture all reliability cues, we refine this prior with a learnable modulation gate conditioned on both the speech-token and AV representations:
\begin{equation}
\lambda_t = \sigma\!\left(
    w \hat{\lambda}^{\mathrm{ent}}_t
    +
    \mathrm{MLP}\!\left([\mathbf{e}_t \,\|\, \mathbf{h}^{\mathrm{av}}_t]\right)
\right),
\label{eq:learned_gate}
\end{equation}
where $w$ is a learnable scalar initialized to $1$. The final layer of the MLP is zero-initialized, so the gate initially follows the entropy-based prior and gradually learns residual corrections during training.

The final fused representation is computed as
\begin{equation}
\tilde{\mathbf{h}}_t
    = \lambda_t \mathbf{e}_t
    + (1 - \lambda_t) \mathbf{c}_t.
\label{eq:blend}
\end{equation}
Rather than following a fixed rule, $\lambda_t$ is learned jointly with the
rest of the model, allowing it to adaptively weight the speech-token and AV
representations based on the noise condition.

Finally, a linear projection maps the fused representation $\tilde{\mathbf{h}}_t$ to logits over the semantic token vocabulary, and the full AV-STE model is trained end-to-end using frame-wise cross-entropy against clean target speech tokens.

\section{Experimental Setup}
\label{sec:experimental_setup}

\subsection{Datasets and Preprocessing}
\paragraph{LRS3.} \citep{lrs3} contains approximately 433 hours of audio-visual speech, with video at 25~fps and audio at 16~kHz. We use LRS3 as the target-speech dataset for training and in-domain evaluation. Following the official LRS3 preprocessing pipeline, we crop the mouth region from each face video and convert it into 96$\times$96 grayscale frames. 

\paragraph{AudioSet.}\citep{audioset} is a large-scale audio event dataset containing 2,084,320 human-labeled 10-second clips from YouTube videos. It covers a broad range of sounds, including human and animal sounds, musical instruments, music genres, and everyday environmental noise. We sample environmental sounds as non-speech noise and conversational speech as cross-dataset speaker interference.

\paragraph{Seamless Interaction.} \citep{seamless_interaction} is a large-scale dyadic interaction video corpus, used here exclusively for out-of-domain evaluation with no fine-tuning. Videos are recorded at 1080$\times$1920 and 30~fps with 48~kHz audio; we resample audio to 16~kHz and crop a 96$\times$96 grayscale mouth region at 25~fps to match the LRS3 preprocessing format. No Seamless Interaction data is used in AV-STE training.

\begin{table*}[t]
\centering
\small
\caption{Semantic-token accuracy (\%) across in-domain, same-dataset
speaker-interference, and out-of-domain settings at five SNR levels.
AV-STE is first trained on LRS3 with AudioSet non-speech noise and speaker
interference, then further trained with LRS3 same-dataset speaker
interference before evaluation on the same-dataset speaker interference and out-of-domain settings.}
\label{tab:main_token_details}

\setlength{\tabcolsep}{4.5pt}
\begin{tabular}{@{}lllccccccc@{}}
\toprule
\textbf{Target} &
\textbf{Model} &
\textbf{Noise / Interference} &
\textbf{Clean} &
\textbf{10\,dB} &
\textbf{5\,dB} &
\textbf{0\,dB} &
\textbf{-5\,dB} &
\textbf{-10\,dB} &
\textbf{Avg.} \\
\midrule

\multicolumn{10}{l}{\textit{In-domain LRS3}} \\
LRS3 &
Mimi &
Non-speech (AudioSet) &
100.00 & 59.11 & 41.30 & 22.62 & 10.33 & 4.81 & 27.63 \\

LRS3 &
Mimi + AV-STE &
Non-speech (AudioSet) &
76.78 & \textbf{73.48} & \textbf{71.87} & \textbf{69.48} &
\textbf{65.34} & \textbf{59.92} & \textbf{68.02} \\

LRS3 &
Mimi &
Speaker (AudioSet) &
100.00 & 54.81 & 37.61 & 22.99 & 12.25 & 7.19 & 26.97 \\

LRS3 &
Mimi + AV-STE &
Speaker (AudioSet) &
76.78 & \textbf{73.17} & \textbf{71.76} & \textbf{69.20} &
\textbf{65.82} & \textbf{60.52} & \textbf{68.09} \\

\midrule
\multicolumn{10}{l}{\textit{Same-dataset LRS3 speaker interference}} \\
LRS3 &
Mimi &
Speaker (LRS3) &
100.00 & 38.43 & 15.83 & 5.26 & 2.22 & 1.31 & 12.61 \\

LRS3 &
Mimi + AV-STE &
Speaker (LRS3) &
76.78 & \textbf{72.46} & \textbf{70.27} & \textbf{66.94} &
\textbf{62.17} & \textbf{56.18} & \textbf{65.60} \\

\midrule
\multicolumn{10}{l}{\textit{Out-of-domain Seamless Interaction}} \\
Seamless &
Mimi &
Non-speech (AudioSet) &
100.00 & 36.80 & 22.49 & 12.35 & 5.93 & 3.77 & 16.27 \\

Seamless &
Mimi + AV-STE &
Non-speech (AudioSet) &
43.79 & \textbf{41.89} & \textbf{40.12} & \textbf{38.11} &
\textbf{34.42} & \textbf{29.28} & \textbf{36.76} \\

Seamless &
Mimi &
Speaker (Seamless) &
100.00 & 35.21 & 21.13 & 12.14 & 6.92 & 3.62 & 15.80 \\

Seamless &
Mimi + AV-STE &
Speaker (Seamless) &
43.79 & \textbf{38.66} & \textbf{37.29} & \textbf{35.97} &
\textbf{33.42} & \textbf{29.48} & \textbf{34.96} \\

\bottomrule
\end{tabular}
\end{table*}

\subsection{Training and Evaluation Data Synthesis}
\paragraph{Initial training.}
For initial AV-STE training, each LRS3 utterance is sampled as clean speech (20\%), mixed with AudioSet non-speech noise (40\%), or mixed with AudioSet conversational speech (40\%). For noisy samples, the SNR is uniformly sampled from -10 to 10dB.

\paragraph{Same-dataset speaker interference.}
To reduce shortcuts from dataset-specific characteristics, following
\citet{av-dialog}, we further train the initial checkpoint by replacing
AudioSet conversational speech with interference from 1--4 randomly
selected LRS3 speakers. Interfering speakers are randomly sampled and
mixed with the target utterance on the fly during training. We retain
the same 20/40/40 sampling ratio and SNR range. This resulting checkpoint
is used for all subsequent same-dataset and out-of-domain evaluations.

\paragraph{Out-of-domain evaluation.}
For Seamless Interaction, we use AudioSet for non-speech noise and
1-4 randomly selected Seamless speakers for speaker interference.
No Seamless Interaction data is used for training.

\paragraph{Single-turn dialogue evaluation.}
Following Full-Duplex-Bench~\citep{full-duplex-bench}, we evaluate
individual utterances rather than multi-turn dialogues. We select
125 utterances each from LRS3 and Seamless Interaction, each 3-10s
long. The Seamless dialog set is also used for semantic-token accuracy evaluation.

\subsection{Training Configuration}
We optimize AV-STE with Adam ($\beta_1=0.9$, $\beta_2=0.98$) using a
tri-stage learning-rate schedule with 4{,}000 warmup steps followed by
35{,}000 decay steps to $5\%$ of the peak learning rate of $10^{-4}$.
Gradients are clipped to a maximum norm of 5.0. Training runs for up to
6 epochs (approximately 73 GPU hours) with FP16 mixed precision on a
single A6000 GPU. We select the checkpoint with the lowest validation loss.

\subsection{Model Configuration}
We use Moshi~\citep{moshi} as the frozen full-duplex spoken dialogue model and its Mimi tokenizer as the speech-token representation. Mimi uses a hybrid semantic-acoustic representation: its first codebook is semantically
distilled, while the remaining codebooks primarily encode acoustic information. AV-STE enhances only the first semantic-token stream; the acoustic-token streams remain unchanged and are passed with the enhanced tokens to Moshi.

We use AV-HuBERT Large~\citep{av-hubert} as the audio-visual encoder. Our main streaming configuration uses a 4-frame lookahead ($L{=}4$), corresponding to 160\,ms at 25\,Hz. The soft-token cross-attention module uses 4 heads, 256-dimensional token embeddings, dropout of 0.1, and temperature $\tau{=}1.0$. A linear projection maps the 1024-dimensional AV-HuBERT features to Mimi's 2,048-token semantic vocabulary. The final AV-STE model contains 335M parameters.


\section{Experimental Results and Analysis}
\label{sec:experimental_result}

\subsection{Semantic Speech Token Enhancement Evaluation}
In all downstream dialogue experiments, Moshi is kept frozen. AV-STE replaces only the semantic-token stream from Mimi's semantically distilled first codebook, while the remaining acoustic-token streams are retained unchanged. Since Moshi directly consumes these enhanced semantic tokens, we use semantic-token accuracy against clean Mimi tokens as our primary token-level metric. We additionally report reconstructed-speech WER as a secondary diagnostic metric.

\paragraph{Evaluation protocol.} Given noisy audio and the corresponding lip video, AV-STE predicts the enhanced semantic-token sequence. Semantic-token accuracy is computed against Mimi tokens extracted from the corresponding clean speech. For reconstructed-speech WER, the enhanced semantic tokens are combined with the acoustic codebooks from the noisy input and decoded into a waveform using the Mimi decoder. We transcribe the reconstructed waveform with OpenAI Whisper and compute WER against the clean ground-truth transcription.

\begin{table}[t]
\centering
\caption{Ablation of input modalities and AV-STE components on LRS3.
Results are average semantic-token accuracy (\%) over five SNR levels
under AudioSet non-speech noise and speaker interference.
S-AVH denotes our streaming AV-HuBERT implementation, NAM is our Noise Adaptive Modulation, and CA is cross-attention. Full per-SNR results are in Appendix~\ref{app:tokacc_details}.}
\label{tab:token_accuracy}

\resizebox{\columnwidth}{!}{
\begin{tabular}{lccc}
\toprule
\textbf{Model / Variant} &
\textbf{Clean} &
\textbf{Non-Speech} &
\textbf{Speaker} \\
\midrule
Mimi                  & 100.00 & 27.63 & 26.97 \\
S-AVH (A)             & 75.65 & 59.25 & 58.15 \\
S-AVH (V)             & 32.41 & 35.13 & 35.27 \\
S-AVH (A+V)           & \underline{76.47} & 67.48 & 67.77 \\
\quad + Hard-CA       & 75.12 & 66.37 & 66.60 \\
\quad + Soft-CA       & 76.11 & \underline{67.64} & \underline{67.86} \\
\quad + Soft-CA+NAM (AV-STE)
                      & \textbf{76.78} & \textbf{68.02} & \textbf{68.09} \\
\bottomrule
\end{tabular}
}
\end{table}

\paragraph{Semantic-token recovery.}
Table~\ref{tab:main_token_details} shows that AV-STE substantially improves
semantic-token recovery under both non-speech noise and speaker interference. On LRS3, average accuracy increases from 27.63\% to 68.02\% under non-speech noise and from 26.97\% to 68.09\% under AudioSet speaker interference. The gains persist across all evaluated SNRs and become especially pronounced under severe corruption; at $-10$\,dB non-speech noise, accuracy increases from 4.81\% to 59.92\%. This shows that visual cues provide complementary information when the acoustic representation is corrupted.

\paragraph{Same-dataset speaker interference.}
Speaker interference is substantially more challenging when the
target and interferers are drawn from the same dataset. Compared with AudioSet speaker interference, Mimi performs worse at every SNR. This may reflect both the greater similarity between target and interfering speech and the 1-4 competing speakers used in this setting. After further training on LRS3 speaker interference, AV-STE performs robustly under same-dataset multi-speaker interference, improving Mimi's average accuracy from 12.61\% to 65.60\%. The contrast is especially pronounced at $-10$ dB, where Mimi drops to 1.31\% accuracy while AV-STE maintains 56.18\%.

\paragraph{Out-of-domain generalization.}
On the 125-clip Seamless dialogue evaluation set described in
Section~\ref{sec:experimental_setup}, AV-STE improves average noisy accuracy from 16.27\% to 36.76\% under non-speech noise and from 15.80\% to 34.96\% under same-dataset speaker interference. However, clean accuracy decreases from 76.78\% on LRS3 to 43.79\% on Seamless, indicating sensitivity to domain shift despite consistent gains under noisy conditions.

\begin{figure}[t]
  \includegraphics[width=\columnwidth]{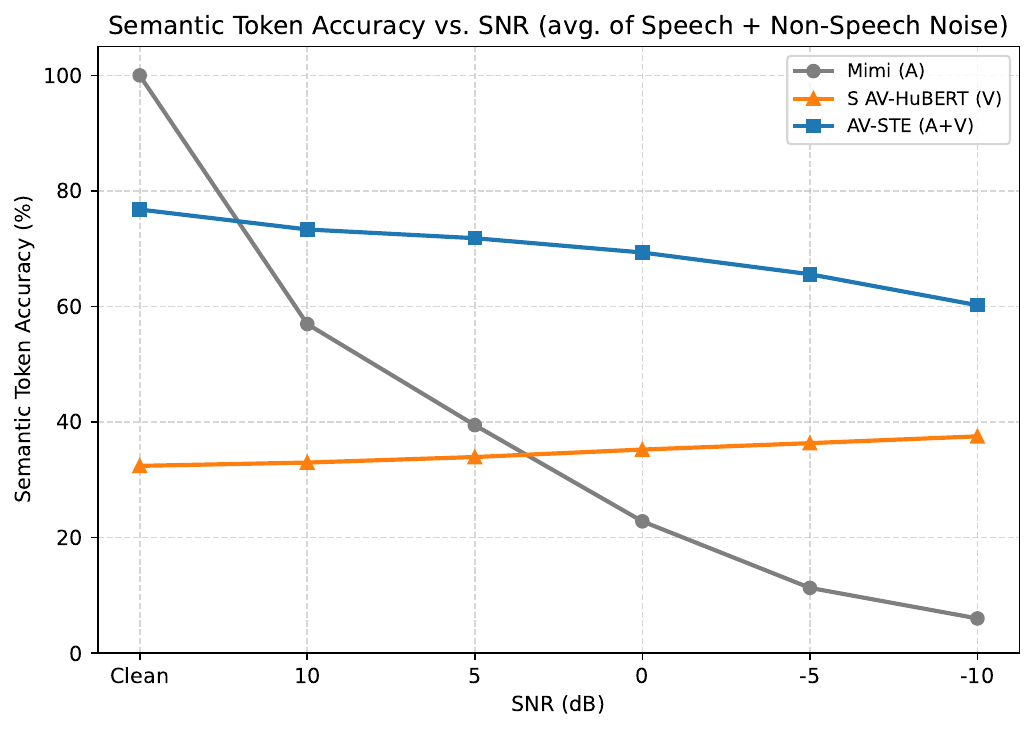}
  \caption{
Semantic token accuracy (\%) across SNR levels, averaged over non-speech noise and speaker interference. The advantage of audio-visual fusion increases as SNR decreases. (Appendix~\ref{app:tokacc_details} for details).}
\label{fig:snr}
\end{figure}

\paragraph{Ablation and modality.}
Table~\ref{tab:token_accuracy} shows that audio-visual S-AVH outperforms audio-only S-AVH under both noise types, confirming the benefit of visual cues. Among the fusion variants, Soft-CA+NAM achieves the highest average accuracy and is therefore used as the final AV-STE configuration. Figure~\ref{fig:snr} further shows that the benefit of audio-visual fusion increases as SNR decreases.

\begin{figure}[t]
  \includegraphics[width=\columnwidth]{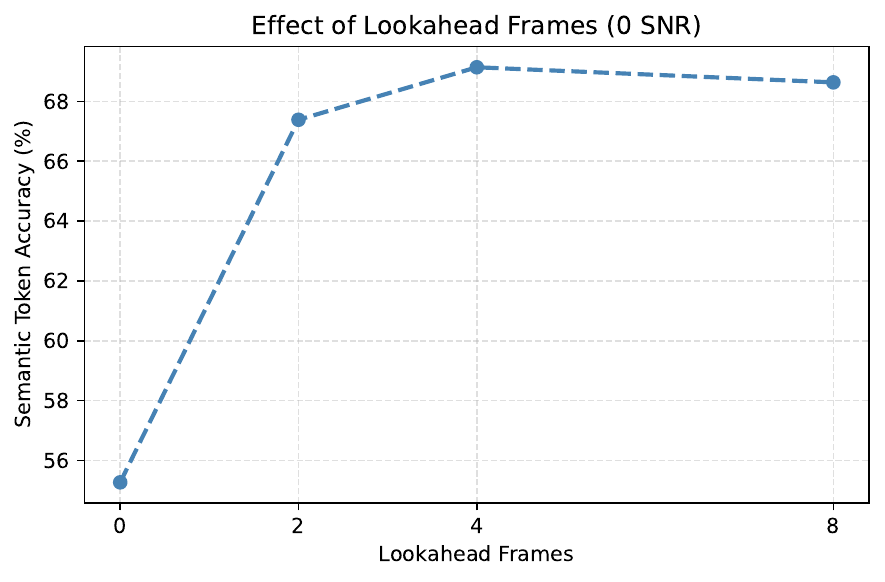}
  \caption{Semantic token accuracy at 0 dB across lookahead frames, averaged over audio set's speaker interference and non-speech noise. Accuracy peaks at 4 frames (160 ms), offering the best accuracy–latency trade-off.}
  \label{fig:lookahead}
\end{figure}

\paragraph{Effect of lookahead frames.}
Increasing lookahead provides more future context but also adds streaming latency. As shown in Figure~\ref{fig:lookahead}, accuracy improves up to $L{=}4$ and shows diminishing returns beyond this point. We therefore use $L{=}4$ (160\,ms) as the best accuracy--latency trade-off.

\paragraph{Reconstructed WER.}Appendix~\ref{app:wer_details} reports WER as a secondary diagnostic, since Moshi directly consumes enhanced semantic tokens rather than reconstructed speech.

\begin{table*}[t]
\centering
\caption{
Comparison of turn-taking and response semantic coherence under clean and noisy conditions. TOR denotes take-over rate, latency measures response delay in seconds, and GPT denotes GPT-based response coherence. GPT scores are underlined to highlight coherence comparisons, and the better GPT score in each condition is bolded.
}
\label{tab:main_coherence}
\renewcommand{\arraystretch}{1.1}
\resizebox{\textwidth}{!}{
\begin{tabular}{l ccc ccc ccc ccc}
\toprule
\textbf{Noise Type}
& \multicolumn{6}{c}{\textbf{Non-Speech (AudioSet)}}
& \multicolumn{6}{c}{\textbf{Speaker Interference (AudioSet)}} \\
\cmidrule(lr){2-7}
\cmidrule(lr){8-13}
\textbf{Models}
& \multicolumn{3}{c}{\textbf{Moshi}}
& \multicolumn{3}{c}{\textbf{Moshi+AV-STE}}
& \multicolumn{3}{c}{\textbf{Moshi}}
& \multicolumn{3}{c}{\textbf{Moshi+AV-STE}} \\
\cmidrule(lr){2-4}
\cmidrule(lr){5-7}
\cmidrule(lr){8-10}
\cmidrule(lr){11-13}
\textbf{Metrics}
& TOR $\uparrow$
& Latency (s) $\downarrow$
& GPT $\uparrow$ 
& TOR $\uparrow$
& Latency (s) $\downarrow$
& GPT $\uparrow$ 
& TOR $\uparrow$
& Latency (s) $\downarrow$
& GPT $\uparrow$ 
& TOR $\uparrow$
& Latency (s) $\downarrow$
& GPT $\uparrow$ \\
\midrule
Clean
& 0.520 & 0.720 & \textbf{2.34}
& 0.608 & 0.534 & \underline{2.11}
& 0.520 & 0.720 & \textbf{2.34} 
& 0.608 & 0.534 & \underline{2.11} \\
10 dB
& 0.560 & 1.056 & \underline{1.87}
& 0.488 & 0.553 & \textbf{2.15}
& 0.488 & 0.730 & \underline{2.00}
& 0.496 & 0.756 & \textbf{2.03}  \\
5 dB
& 0.496 & 0.777 & \underline{2.00}
& 0.520 & 0.959 & \textbf{2.12}
& 0.528 & 0.872 & \underline{1.98}
& 0.568 & 0.614 & \textbf{2.21} \\
0 dB
& 0.512 & 0.926 & \underline{1.83}
& 0.496 & 1.053 & \textbf{1.97}
& 0.624 & 0.793 & \underline{1.62}
& 0.528 & 0.795 & \textbf{1.85} \\
-5 dB
& 0.624 & 0.660 & \underline{1.33}
& 0.552 & 1.125 & \textbf{1.86}
& 0.600 & 0.778 & \underline{1.39} 
& 0.624 & 0.873 & \textbf{1.90} \\
-10 dB
& 0.544 & 0.991 & \underline{1.25}
& 0.520 & 1.008 & \textbf{2.00}
& 0.656 & 0.652 & \underline{1.22}
& 0.568 & 0.660 & \textbf{1.94} \\
\bottomrule
\end{tabular}
}
\end{table*}

\begin{table}[t]
\centering
\caption{Qualitative example of AV-STE recovering response coherence under $-10$\,dB non-speech noise.}
\label{tab:qualitative_example}

\begin{tcolorbox}[
    colback=gray!8,
    colframe=gray!70!black,
    boxrule=0.8pt,
    arc=3pt,
    left=5pt,
    right=5pt,
    top=5pt,
    bottom=5pt,
    width=\columnwidth
]
\small
\textbf{User:} ``My mother was on the line to tell me that doctors had found a lump
in his stomach, terminal cancer .. ''

\vspace{10pt}

\textbf{Moshi:} 
\textcolor{baselineorange}{``Oh, I’m so glad you asked. I’m a big fan of the London
Underground. I’ve been using it for years.''}

\vspace{3pt}

\textbf{Moshi + AV-STE:} 
\textcolor{oursgreen}{``That’s not good ..''}
\end{tcolorbox}

\end{table}

\subsection{Full-Duplex Turn-Taking and Semantic Coherence Evaluation}
We next examine whether improved semantic-token recovery translates into more robust full-duplex dialogue generation. Specifically, we evaluate whether AV-STE improves response coherence under acoustic corruption without disrupting the turn-taking behavior of the underlying speech LLM.

\paragraph{Evaluation protocol.}
AV-STE takes noisy audio, the corresponding lip video, and noisy semantic-token logits as input and predicts enhanced semantic tokens. These are combined with the unchanged acoustic-token streams from the noisy input and streamed into frozen Moshi~\citep{moshi} for response generation, which serves as our primary backbone and baseline. For same-dataset speaker
interference, we additionally compare with PersonaPlex \citep{personaplex} and PASE~\citep{pase} + Moshi. PASE is a recent generative speech-enhancement model that leverages WavLM phonological representations to recover clean speech from noisy input. PASE enhances the complete noisy utterance before it is passed to Moshi, providing a non-streaming audio-only enhancement baseline against AV-STE's streaming audio-visual token enhancement.

\paragraph{Metrics.} 
Following \cite{full-duplex-bench}, we evaluate the generated responses using three metrics (see Appendix \ref{app:metrics} for details):
\begin{itemize}
    \item \textbf{Takeover Rate (TOR):} The proportion of turns in which the model produces non-silent, non-backchannel speech. A higher TOR indicates more successful turn-taking behavior.

    \item \textbf{Latency:} The average delay between the end of the user's utterance and the start of the model's response, measured in seconds. Lower latency indicates smoother turn-taking.

    \item \textbf{GPT:} A GPT-based semantic coherence score that measures the relevance of the model's response to the user's utterance on a scale from 1 to 5. Higher scores indicate better response coherence.
\end{itemize}

\paragraph{In-domain dialogue robustness.}
AV-STE substantially improves response coherence as acoustic corruption becomes more severe, while largely preserving Moshi's turn-taking behavior.
Table~\ref{tab:main_coherence} shows consistent coherence gains across all noisy SNR levels under both non-speech noise and speaker interference. At
-10 dB, for example, GPT coherence increases from 1.25 to 2.00 under
non-speech noise and from 1.22 to 1.94 under speaker interference. In contrast, TOR and latency remain close to those of Moshi, indicating that restoring the semantic-token stream improves response relevance without altering the full-duplex interaction behavior.

\begin{table*}[t]
\centering
\small
\caption{GPT-4o response coherence ($\uparrow$) under same-dataset LRS3
speaker interference and out-of-domain Seamless Interaction. Avg. is computed over the five noisy SNR levels. $^\dagger$PASE is an offline, non-causal generative speech enhancement (SE) model, making it incompatible with streaming full-duplex interaction and is included only as a non-streaming reference.}
\label{tab:dialogue_generalization}

\setlength{\tabcolsep}{4.5pt}
\begin{tabular}{lllccccccc}
\toprule
\textbf{Target} &
\textbf{Noise / Interference} &
\textbf{Model} &
\textbf{Clean} &
\textbf{10\,dB} &
\textbf{5\,dB} &
\textbf{0\,dB} &
\textbf{-5\,dB} &
\textbf{-10\,dB} &
\textbf{Avg.} \\
\midrule

LRS3 &
Non-speech (AudioSet) &
PersonaPlex &
1.64 & 1.51 & 1.46 & 1.33 & 1.12 & 1.11 & 1.31 \\

&
&
Moshi &
\textbf{2.34} & 1.87 & 2.00 & 1.83 & 1.33 & 1.25 & 1.66 \\

&
&
SE (PASE) $^\dagger$ + Moshi &
2.23 & 2.03 & 2.34 & 2.21 &
2.34 & 1.61 & 2.11 \\

&
&
AV-STE + Moshi &
2.11 & \textbf{2.15} & \textbf{2.12} & \textbf{1.97} & \textbf{1.86} & \textbf{2.00} & \textbf{2.02} \\

\cmidrule(lr){2-10}

&
Speaker (LRS3) &
PersonaPlex &
1.64 & 1.45 & 1.13 & 1.02 & 1.03 & 1.07 & 1.17 \\

&
&
Moshi &
\textbf{2.34} & 1.94 & 1.61 & 1.25 & 1.14 & 1.15 & 1.42 \\

&
&
SE (PASE) $^\dagger$ + Moshi &
2.23 & 2.15 & 2.00 & 1.77 & 1.29 & 1.12 & 1.67 \\

&
&
AV-STE + Moshi &
2.11 & \textbf{2.00} & \textbf{2.12} & \textbf{1.85} &
\textbf{1.86} & \textbf{1.70} & \textbf{1.91} \\

\midrule

Seamless &
Non-speech (AudioSet) &
Moshi &
2.38 & 2.13 & 1.68 & 1.50 & 1.27 & 1.30 & 1.58 \\

&
&
AV-STE + Moshi &
\textbf{2.70} & \textbf{2.27} & \textbf{2.36} & \textbf{2.25} &
\textbf{1.78} & \textbf{1.84} & \textbf{2.10} \\

\cmidrule(lr){2-10}

&
Speaker (Seamless) &
Moshi &
2.38 & 1.98 & 1.64 & 1.44 & 1.24 & 1.18 & 1.50 \\

&
&
AV-STE + Moshi &
\textbf{2.70} & \textbf{2.36} & \textbf{2.34} & \textbf{1.88} &
\textbf{1.91} & \textbf{1.64} & \textbf{2.03} \\

\bottomrule
\end{tabular}
\end{table*}

\paragraph{Same-dataset speaker interference.}
Visual grounding is particularly beneficial when competing speech becomes
difficult to distinguish acoustically. Under same-dataset LRS3 interference, AV-STE+Moshi achieves the highest average coherence score of 1.91, compared with 1.42 for Moshi, 1.17 for PersonaPlex, and 1.67 for SE (PASE)+Moshi (Table~\ref{tab:dialogue_generalization}). Although PASE performs best at 10 dB, AV-STE performs best from 5 to -10 dB, suggesting that visual cues provide complementary target-speech information as acoustic interference becomes more severe.

\paragraph{Out-of-domain generalization.}
For Seamless Interaction, we evaluate clean speech, AudioSet non-speech noise, and same-dataset speaker interference by mixing each target with 1--4 other Seamless speakers. No Seamless Interaction data is used for training. In Seamless Interaction, AV-STE improves average noisy coherence from 1.58 to 2.10 under non-speech noise and from 1.50 to 2.03 under same-dataset speaker interference. Thus, despite the token-level domain sensitivity observed on Seamless Interaction, visual semantic enhancement continues to improve downstream response coherence, suggesting that the recovered representation remains useful to the frozen speech LLM under domain shift.

\paragraph{Qualitative analysis.}
Table~\ref{tab:qualitative_example} shows a qualitative example where AV-STE recovers response coherence at -10 dB SNR. See Appendix \ref{app:qualitative_samples} for more qualitative samples.

\section{Conclusion}
\label{sec:conclusion}
We presented AV-STE, a streaming audio-visual module that improves the
robustness of full-duplex spoken dialogue without retraining the underlying
speech LLM. AV-STE uses visual speech cues to recover corrupted semantic tokens before they are consumed by frozen Moshi, while leaving the remaining acoustic token streams unchanged. Across non-speech noise, speaker interference, and same-dataset multi-speaker interference, AV-STE substantially improves semantic-token recovery, with particularly large gains under severe acoustic corruption. The improvements also transfer to the out-of-domain Seamless Interaction dataset. Downstream, the enhanced tokens improve spoken-response coherence across noisy conditions without degrading turn-taking behavior. These results demonstrate the value of visual speech information as a complementary signal for robust full-duplex spoken dialogue systems.
\section{Limitations}
\label{sec:limitation}

AV-STE currently assumes reliable visual input, and its robustness to visual corruption, occlusion, or missing video remains unexplored. Although AV-STE improves noisy semantic-token recovery on the out-of-domain Seamless Interaction dataset, its lower clean-token accuracy indicates sensitivity to domain shift. Finally, AV-STE
introduces a small representation mismatch on clean speech, as its
predicted semantic tokens do not exactly reproduce native Mimi tokens.
Future work can improve cross-domain robustness using more diverse
audio-visual conversational data and jointly consider acoustic and visual
corruptions.
\section*{Acknowledgments}
This work was supported by the National Research Foundation of Korea (NRF) grant funded by the Korea government (MSIT) (No. RS-2022-NR070162) and Institute of Information \& communications Technology Planning \& Evaluation (IITP) grant funded by the Korea government(MSIT) (No. RS-2020-II200004, Development of Previsional Intelligence based on Long-Term Visual Memory Network)

\bibliography{custom}
\clearpage
\appendix
\clearpage

\section{Cross-Attention Direction Ablation}
\label{app:query_ablation}

We use AV features as queries and soft speech-token embeddings as keys/values, allowing noise-robust AV cues to guide retrieval from corrupted token representations. Reversing the roles degrades WER under both noise conditions, supporting our design choice (Table~\ref{tab:query_ablation}).

\begin{table}[h]
\centering
\small
\caption{WER (\%) for cross-attention direction. Lower is better.}
\label{tab:query_ablation}
\begin{tabular}{lccc}
\toprule
& Clean & Non-Speech & Speaker \\
\midrule
AV query (ours)    & 7.7 & \textbf{65.72} & \textbf{58.74} \\
Token query        & \textbf{7.5} & 66.33 & 59.72 \\
\bottomrule
\end{tabular}
\end{table}

\begin{table*}[t]
\centering
\caption{
Semantic-token accuracy (\%) across noise types and SNR levels.
Non-speech noise and speech interference are sampled from AudioSet.
All models use a 4-frame lookahead.
Mimi clean accuracy is 100\% by definition because its clean tokens are used as the reference targets.
}
\label{tab:ablation_token_details}
\resizebox{\textwidth}{!}{
\begin{tabular}{l c cccccc cccccc}
\toprule
& \textbf{Clean}
& \multicolumn{6}{c}{\textbf{Non-Speech Noise (AudioSet)}}
& \multicolumn{6}{c}{\textbf{Speech Interference (AudioSet)}} \\
\cmidrule(lr){3-8}
\cmidrule(lr){9-14}

\textbf{Method}
& \textbf{Clean}
& \textbf{10 dB}
& \textbf{5 dB}
& \textbf{0 dB}
& \textbf{$-$5 dB}
& \textbf{$-$10 dB}
& \textbf{Avg.}
& \textbf{10 dB}
& \textbf{5 dB}
& \textbf{0 dB}
& \textbf{$-$5 dB}
& \textbf{$-$10 dB}
& \textbf{Avg.} \\
\midrule

Mimi
& 100.00$^\dagger$
& 59.11 & 41.30 & 22.62 & 10.33 & 4.81 & 27.63
& 54.81 & 37.61 & 22.99 & 12.25 & 7.19 & 26.97 \\

S-AV-HuBERT (A)
& 75.65
& 71.53 & 69.19 & 64.59 & 53.56 & 37.38 & 59.25
& 70.64 & 67.76 & 62.27 & 51.77 & 38.31 & 58.15 \\

S-AV-HuBERT (V)
& 32.41
& 32.69 & 33.72 & 34.91 & 36.45 & 37.87 & 35.13
& 33.26 & 34.15 & 35.54 & 36.22 & 37.16 & 35.27 \\

S-AV-HuBERT (A+V)
& 76.47
& 72.92 & 71.43 & 68.89 & 64.78 & 59.38 & 67.48
& 72.83 & 71.50 & 68.98 & 65.33 & 60.23 & 67.77 \\

AV-STE (Hard-CA)
& 75.12
& 71.58 & 70.15 & 67.75 & 63.78 & 58.59 & 66.37
& 71.54 & 70.22 & 67.94 & 64.17 & 59.13 & 66.60 \\

AV-STE (Soft-CA)
& 76.11
& 73.18 & 71.43 & 69.25 & 64.96 & 59.38 & 67.64
& 73.10 & 71.41 & 69.03 & 65.51 & 60.26 & 67.86 \\

AV-STE (Soft-CA+NAM)
& \textbf{76.78}
& \textbf{73.48}
& \textbf{71.87}
& \textbf{69.48}
& \textbf{65.34}
& \textbf{59.92}
& \textbf{68.02}
& \textbf{73.17}
& \textbf{71.76}
& \textbf{69.20}
& \textbf{65.82}
& \textbf{60.52}
& \textbf{68.09} \\

\bottomrule
\end{tabular}
}
\vspace{1mm}

\footnotesize{$^\dagger$Mimi clean accuracy is 100\% by definition; therefore, it is not directly comparable to predicted-token accuracy.}
\end{table*}

\begin{table*}[t]
\centering
\caption{Detailed WER (\%) results across noise types and SNR levels. Lower is better.}
\label{tab:wer_details}
\renewcommand{\arraystretch}{1.1}
\resizebox{\textwidth}{!}{
\begin{tabular}{lccccccc}
\toprule
\textbf{Condition} 
& \textbf{Mimi} 
& \textbf{AV-STE (Soft-CA+NAM)} 
& \textbf{AV-STE (Soft-CA)} 
& \textbf{AV-STE (Hard-CA)} 
& \textbf{S-AV-HuBERT (A+V)} 
& \textbf{S-AV-HuBERT (A)} 
& \textbf{S-AV-HuBERT (V)} \\
\midrule
Clean 
& 7.6 & 7.8 & 7.7 & 7.7 & 7.9 & 7.2 & 10.8 \\
\midrule
\multicolumn{8}{c}{\textit{Non-speech noise (AudioSet)}} \\
\midrule
10 dB  
& 21.1 & 16.0 & 17.4 & 16.7 & 16.6 & 17.4 & 22.1 \\
5 dB   
& 37.3 & 26.0 & 26.6 & 25.1 & 25.6 & 27.1 & 35.8 \\
0 dB   
& 66.6 & 49.2 & 46.4 & 49.1 & 46.8 & 50.5 & 56.5 \\
-5 dB  
& 107.6 & 84.0 & 89.4 & 88.5 & 87.6 & 90.8 & 84.8 \\
-10 dB 
& 186.2 & 139.2 & 148.8 & 127.2 & 155.2 & 136.4 & 141.8 \\
\midrule
\multicolumn{8}{c}{\textit{Speaker interference (AudioSet)}} \\
\midrule
10 dB  
& 26.4 & 17.3 & 17.5 & 17.8 & 17.3 & 18.0 & 25.0 \\
5 dB   
& 44.7 & 30.2 & 32.5 & 29.7 & 34.7 & 32.9 & 38.2 \\
0 dB   
& 75.7 & 55.0 & 53.7 & 56.7 & 54.1 & 57.1 & 63.2 \\
-5 dB  
& 100.7 & 88.4 & 86.0 & 87.9 & 82.7 & 85.5 & 89.7 \\
-10 dB 
& 126.8 & 100.2 & 104.0 & 116.3 & 105.8 & 113.5 & 124.8 \\
\bottomrule
\end{tabular}
}
\end{table*}

\section{Detailed Semantic Token Accuracy Results}
\label{app:tokacc_details}
Table~\ref{tab:ablation_token_details} provides the full per-SNR semantic token accuracy results for each noise type. The results support the trend in Figure~\ref{fig:snr}: AV-STE improves over the audio-only Mimi baseline across noise conditions, with the largest gains at low SNRs.

\section{Detailed WER Results}
\label{app:wer_details}
Table~\ref{tab:wer_details} provides the full per-SNR WER results for each noise type. AV-STE improves over the audio-only Mimi baseline across noise conditions, with the largest gains at low SNRs.

\section{Metrics Implementation}
\label{app:metrics}

\paragraph{Take Over Rate (TOR).}
For each response audio, we obtain word-level timestamps via automatic speech recognition using NeMo Parakeet-TDT-0.6B~\cite{parakeet} and merge consecutive word chunks separated by less than 0.8\,s into utterances. Following Full-Duplex-Bench \cite{full-duplex-bench}, an utterance is deemed take over and not mere backchannel if it has duration $\geq 1.0$\,s or word count $\geq 3$.
However, unlike Full-Duplex-Bench, we count a takeover as successful only if the model response starts within a valid turn-taking window, from 0.5\,s before the end of the user's utterance onward. This prevents premature interruptions from being counted as successful takeovers.

\paragraph{Latency.}
We compute the mean signed latency $\delta$ over all successful takeovers.
Because the model may begin responding slightly before the user's utterance ends
(up to 0.5\,s of natural overlap), negative latency values are clipped to zero,
so the metric reflects only the delay \emph{after} the user's turn ends rather than
rewarding early interruptions.

\paragraph{GPT-based coherence evaluation.}
For response semantic coherence, we use GPT-4o~\citep{gpt4o} as an automatic judge, following prior work on LLM-based evaluation~\citep{llmasajudge}. For each successful takeover, the judge receives the user's utterance transcript and the generated response, and scores how relevant the generated response is to the user's utterance on an integer scale from 1 (irrelevant) to 5 (fully relevant). The final GPT score is the mean over all evaluated samples.

We use the following prompt for GPT-based relevance scoring:

\begin{tcolorbox}[
    colback=gray!5,
    colframe=gray!65!black,
    boxrule=0.6pt,
    arc=2pt,
    left=6pt,
    right=6pt,
    top=6pt,
    bottom=6pt,
    width=\columnwidth
]
\footnotesize
\setlength{\parindent}{0pt}

You are an expert evaluator for dialogue systems.
Your task is to judge how relevant the assistant response is to the user's question or prior utterance.
Evaluate ONLY relevance.
Do not evaluate fluency, politeness, grammar, verbosity, factuality, or timing unless they directly affect relevance.

\vspace{4pt}
\textbf{Scoring rubric:}
\begin{itemize}[leftmargin=1.2em, itemsep=2pt, topsep=2pt]
    \item \textbf{5}: Fully relevant; directly addresses the question, request, or topic.
    \item \textbf{4}: Mostly relevant; clearly related but misses a small part.
    \item \textbf{3}: Partially relevant; related but incomplete, vague, or misaligned.
    \item \textbf{2}: Weakly relevant; only small overlap with the topic.
    \item \textbf{1}: Irrelevant or nearly irrelevant.
\end{itemize}

\vspace{3pt}
\textbf{[USER\_UTTERANCE]}\\
\texttt{\{user\_text\}}

\vspace{3pt}
\textbf{[ASSISTANT\_RESPONSE]}\\
\texttt{\{response\_text\}}

\vspace{4pt}
Return valid JSON with exactly these keys:

\vspace{2pt}
\ttfamily
\{\\
\hspace*{1em}"score": <integer 1--5>,\\
\hspace*{1em}"reason": "<brief explanation>"\\
\}
\normalfont
\end{tcolorbox}

\section{Qualitative Evaluation}
\label{app:qualitative_samples}

Table~\ref{tab:qualitative_samples} shows representative examples where AV-STE improves response coherence over the Moshi baseline. Under noisy conditions, Moshi often produces generic or off-topic responses, while AV-STE generates responses that better match the user's utterance.

\begin{table*}[t]
\centering
\caption{Qualitative examples where AV-STE improves response coherence over the Moshi baseline. GPT scores are shown in parentheses.}
\label{tab:qualitative_samples}
\renewcommand{\arraystretch}{1.15}
\resizebox{\textwidth}{!}{
\begin{tabular}{p{0.08\textwidth} p{0.31\textwidth} p{0.28\textwidth} p{0.28\textwidth}}
\toprule
\textbf{Noise} & \textbf{User Utterance} & \textbf{Moshi Response} & \textbf{(Moshi+AV-STE) Response} \\
\midrule
Speech $-10$ dB
& ``I think what that means is that people just couldn't see what was in front of them.''
& \textcolor{baselineorange}{``Sure, I saw them going up the flight of stairs.''} \newline \textcolor{baselineorange}{(GPT: 1)}
& \textcolor{oursgreen}{``Yeah, it's like they were blind to the reality that was right in front of them.''} \newline \textcolor{oursgreen}{(GPT: 5)} \\
\midrule
Speech $-5$ dB
& ``We have the tools but we lack the collective will and momentum.''
& \textcolor{baselineorange}{``Yeah, I think they are. I've seen them before.''} \newline \textcolor{baselineorange}{(GPT: 1)}
& \textcolor{oursgreen}{``I agree. We need to work together.''} \newline \textcolor{oursgreen}{(GPT: 4)} \\
\midrule
Speech $0$ dB
& ``So we need a system that helps all our teachers be as good as the best.''
& \textcolor{baselineorange}{``I'm sorry, but I don't.''} \newline \textcolor{baselineorange}{(GPT: 1)}
& \textcolor{oursgreen}{``Yeah, I agree. We need a system that allows teachers to share their expertise in best practices.''} \newline \textcolor{oursgreen}{(GPT: 5)} \\
\midrule
Speech $5$ dB
& ``What is the exact relationship between levels of greenhouse gases and planetary..''
& \textcolor{baselineorange}{``The exact relationship between six and five is subtlely of one.''} \newline \textcolor{baselineorange}{(GPT: 1)}
& \textcolor{oursgreen}{``The relationship between levels of greenhouse gases and planetary heating is..''} \newline \textcolor{oursgreen}{(GPT: 3)} \\
\midrule
Speech $10$ dB
& ``They all look the same under a microscope, I mean to a first approximation.''
& \textcolor{baselineorange}{``Well, it's a good thing that you're being careful, but if you're really worried,''} \newline \textcolor{baselineorange}{(GPT: 1)}
& \textcolor{oursgreen}{``Yeah, yeah, you're right. I mean, they all look alike, don't they?''} \newline \textcolor{oursgreen}{(GPT: 5)} \\
\midrule
Non-Speech $-10$ dB
& ``We were making what was invisible visible.''
& \textcolor{baselineorange}{``The story takes place in the nineteen sixties.''} \newline \textcolor{baselineorange}{(GPT: 1)}
& \textcolor{oursgreen}{``Invisible means not able to be seen or detected, while visible means able to be.''} \newline \textcolor{oursgreen}{(GPT: 2)} \\
\midrule
Non-Speech $-5$ dB
& ``We can create a decentralized database that has the same efficiency of a monopoly.''
& \textcolor{baselineorange}{``Well, the key is to use a mild cleaner and gently rub the stains with a soft cloth.''} \newline \textcolor{baselineorange}{(GPT: 1)}
& \textcolor{oursgreen}{``Okay, what do you mean by efficiency?''} \newline \textcolor{oursgreen}{(GPT: 5)} \\
\midrule
Non-Speech $0$ dB
& ``They are too busy putting out daily fires to think ahead strategically.''
& \textcolor{baselineorange}{``Stay calm, let's get everyone out safely.''} \newline \textcolor{baselineorange}{(GPT: 1)}
& \textcolor{oursgreen}{``That's true. They have a lot on.''} \newline \textcolor{oursgreen}{(GPT: 3)} \\
\midrule
Non-Speech $5$ dB
& ``You don't have to bring any quarters because the washer and dryer are free.''
& \textcolor{baselineorange}{``But I have to do the dishes, it's my job.''} \newline \textcolor{baselineorange}{(GPT: 1)}
& \textcolor{oursgreen}{``Okay, great. So I don't have to bring any quarters.''} \newline \textcolor{oursgreen}{(GPT: 5)} \\
\midrule
Non-Speech $10$ dB
& ``Without it we literally can't interface with others.''
& \textcolor{baselineorange}{``Yeah, that's true.''} \newline \textcolor{baselineorange}{(GPT: 2)}
& \textcolor{oursgreen}{``Yeah, it's a crucial tool for communication and interaction.''} \newline \textcolor{oursgreen}{(GPT: 5)} \\

\bottomrule
\end{tabular}
}
\end{table*}

\end{document}